\documentclass{aa} % for the letters 
\usepackage{graphicx}
\usepackage{txfonts}
\usepackage{xspace}
\usepackage{natbib}
\usepackage{color}
\newcommand{\Msun}{$M_{\odot}$\xspace}
\newcommand{\Lsun}{$L_{\odot}$\xspace}
\newcommand{\Msunyr}{$M_{\odot}$\,yr$^{-1}$\xspace}

\newcommand{\vlsr}{$\varv_{\rm{LSR}}$\xspace}
\newcommand{\vlsrstar}{$\varv_{\rm{LSR}}^*$}

\newcommand{\kms}{km\,s$^{-1}$\xspace}
\newcommand{\CO}{$^{12}$CO}
\newcommand{\COone}{$^{12}\rm{CO}(1-0)$\xspace}
\newcommand{\COtwo}{$^{12}\rm{CO}(2-1)$\xspace}
\newcommand{\COthree}{$^{12}\rm{CO}(3-2)$\xspace}
\xspace

\begin{document}
   \title{A detailed view of the gas shell around R Sculptoris with ALMA}

%   \subtitle{ALMA observations of R Sculptoris}

   \author{M.~Maercker
          \inst{1}
          \and
          W.H.T.~Vlemmings \inst{1}
          \and
          M.~Brunner \inst{2}
          \and
          E.~De~Beck \inst{1}
          \and
          E.M.~Humphreys \inst{3}
          \and
          F.~Kerschbaum \inst{2}
          \and
          M.~Lindqvist \inst{1}
          \and
          H.~Olofsson \inst{1}
          \and
          S.~Ramstedt \inst{4}
         %\fnmsep\thanks{Just to show the usage of the elements in the author field}
          }

     \institute{Department of Earth and Space Sciences, Chalmers University of Technology, Onsala Space Observatory, 43992 Onsala, Sweden\\
   \email{maercker@chalmers.se}
         \and
             Department of Astrophysics, University of Vienna, T\"urkenschanzstr. 17, 1180 Vienna, Austria
             \and
             European Southern Observatory, Karl-Schwarzschild-Straße 2, 85748, Garching, Germany
             \and
             Department of Physics and Astronomy, Uppsala University, Box 516, 75120 Uppsala, Sweden\\
             %\thanks{The university of heaven temporarily does not accept e-mails}
             }

   \date{Received 5 August 2015 / Accepted 17 November 2015}

% \abstract{}{}{}{}{} 
% 5 {} token are mandatory
 
  \abstract
  % context heading (optional)
  % {} leave it empty if necessary  
   {During the asymptotic giant branch (AGB) phase, stars undergo thermal pulses $-$ short-lived phases of explosive helium burning in a shell around the stellar core. Thermal pulses lead to the formation and mixing-up of new elements to the stellar surface. They are hence fundamental to the chemical evolution of the star and its circumstellar envelope. A further consequence of thermal pulses is the formation of detached shells of gas and dust around the star, several of which have been observed around carbon-rich AGB stars.}
  % aims heading (mandatory)
   {We aim to determine the physical properties of the detached gas shell around R Sculptoris, in particular the shell mass and temperature, and to constrain the evolution of the mass-loss rate during and after a thermal pulse.}
  % methods heading (mandatory)
   {We analyse \COone, \COtwo, and \COthree~emission, observed with the Atacama Large Millimeter/submillimeter Array (ALMA) during Cycle 0 and complemented by single-dish observations. The spatial resolution of the ALMA data allows us to separate the detached shell emission from the extended emission inside the shell. We perform radiative transfer modelling of both components to determine the shell properties and the post-pulse mass-loss properties.}
  % results heading (mandatory)
   {The ALMA data show a gas shell with a radius of 19\farcs5 expanding at 14.3\,\kms. The different scales probed by the ALMA Cycle 0 array show that the shell must be entirely filled with gas, contrary to the idea of a detached shell. The comparison to single-dish spectra and radiative transfer modelling confirms this. We derive a shell mass of 4.5$\times10^{-3}$\,\Msun with a temperature of 50\,K. Typical timescales for thermal pulses imply a pulse mass-loss rate of 2.3$\times10^{-5}$\,\Msunyr. For the post-pulse mass-loss rate, we find evidence for a gradual decline of the mass-loss rate, with an average value of $1.6\times10^{-5}$\,\Msunyr. The total amount of mass lost since the last thermal pulse is 0.03\,\Msun, a factor four higher compared to classical models, with a sharp decline in mass-loss rate immediately after the pulse.}
  % conclusions heading (optional), leave it empty if necessary 
   {We find that the mass-loss rate after a thermal pulse has to decline more slowly than generally expected from models of thermal pulses. This may cause the star to lose significantly more mass during a thermal pulse cycle, which affects the lifetime on the AGB and the chemical evolution of the star, its circumstellar envelope, and the interstellar medium.}


   \maketitle
%
%________________________________________________________________

\section{Introduction}

The chemical evolution of evolved stars between 0.8--10 \Msun on the asymptotic giant branch (AGB) is driven by thermal pulses (TP). Thermal runaway burning of helium in a shell around the dormant carbon--oxygen core leads to a mixing of the inner layers of the star, resulting in an intricate nucleosynthetic network and the production of new heavy elements~\citep[e.g. ][]{karakasco2007}. These elements are mixed to the surface of the star through deep convection during the third dredge-up after the pulse and are incorporated into the stellar wind. The winds from AGB stars replenish the interstellar medium (ISM) with this newly processed material~\citep[e.g. ][]{schneideretal2014}. The stellar mass loss is high enough (up to 10$^{-4}$\,\Msunyr) to eventually terminate the evolution of the star on the AGB with most of the stellar mass lost to the ISM~\citep[e.g. ][]{herwig2005}. As such, AGB stars are significant contributors to the chemical evolution of the ISM and galaxies. The material is included in molecular clouds where new stars and planets are formed. 

The phase of rapid helium burning during a TP cycle lasts only a few hundred years~\citep{vassiliadisco1993}. Although the helium luminosity experiences a large increase, the surface luminosity only increases slightly. However, this increase in luminosity leads to additional radiation pressure on the dust grains in the upper atmosphere of the star and, combined with a lower temperature and larger radius, results in an increased mass-loss rate and expansion velocity. After He-burning ceases, the surface luminosity, and hence the mass-loss rate and expansion velocity, decrease again~\citep{steffenco2000,mattssonetal2007}
	
	An indication of highly variable mass loss from AGB stars was first observed in CO emission in the form of detached shells around carbon AGB stars, and a formation connected to TPs was suggested \citep{olofssonetal1988,olofssonetal1990}. Since then, detached shells of dust and gas around $\sim$ten carbon AGB stars have been observed. The strongest argument for the connection between detached-shell sources and TPs are the detection statistics that are based on a volume-limited sample of carbon stars~\citep{olofssonetal1993a}. No detached CO shells have been observed around oxygen-rich (M-type) AGB stars, where dust opacity effects possibly prevent the formation of a shell during the TP~\citep[e.g. ][]{bladhetal2015}. The formation mechanism is also fundamentally different from the shells observed around high-mass evolved stars, where photoionization creates confined shells~\citep{mackeyetal2014}. Thermal emission from the dust in detached shells was first observed in the far-infrared by IRAS~\citep{vanderveenco1988,watersetal1994,izumiuraetal1997}, and more recently by AKARI and Herschel~\citep{kerschbaumetal2010,izumiuraetal2011,coxetal2012,mecinaetal2014}. Most of these shells are likely connected to TPs. The images make it possible to determine the dust temperature in the detached shells and, to some extent, the sizes of the shell. Observations of dust scattered, stellar light in the optical have given a detailed view of the dust distribution in the shells, most recently in scattered polarised light of the detached shell sources R~Scl and V644~Sco~\citep{delgadoetal2001,delgadoetal2003a,olofssonetal2010,maerckeretal2010,maerckeretal2014,ramstedtetal2011}. In comparison to single-dish (SD) observations of CO line emission and images of dust in the far infrared, the optical observations of dust scattered light provide information at high angular resolution.
Observations at high-angular resolution of the detached gas shells in CO emission have only been done for the detached shell sources TT~Cyg~\citep{olofssonetal2000} and U~Cam~\citep{lindqvistetal1999} using the Plateau de Bure Interferometer, and R~Scl~\citep{maerckeretal2012} using the Atacama Large Millimeter/submillimeter Array (ALMA). The observations clearly show the detached shells around TT~Cyg and U~Cam at 35\arcsec\/ and 7\farcs3, respectively. No emission is detected between the shells and the present-day mass-loss, implying that the mass-loss rate decreased significantly after the TP, leading to the formation of an expanding shell that is not connected to the present-day mass-loss (and hence detached). The recent observations with ALMA during Cycle 0 imaged the detached shell around R~Scl in \COthree emission at high angular resolution ~\citep[$\approx$1\farcs4 beam;][]{maerckeretal2012}, showing a clumpy structure and slight deviations from spherical symmetry in the shell. In addition to the detached shell, the data also show a spiral structure that connects the shell with the present-day mass-loss. The spiral is formed by the interaction of the stellar wind with a previously unknown binary companion, and allows us to determine the evolution of the mass-loss rate and expansion velocity since the last TP. These are the first \emph{observational} constraints of the behaviour of the stellar mass loss during and after a TP. While the evolution generally fits with theoretical predictions, it is clear that the mass-loss rate must decrease significantly slower than predicted by models. These observations already show that the shell around R~Scl is not entirely detached. Comparing the distribution of the gas shell with images of polarised, dust scattered stellar light shows that the dust and gas have an almost identical distribution in the shell, implying a common evolution of the dust and gas shells since the last TP~\citep{maerckeretal2014}. 
	
	\cite{maerckeretal2012} only discuss the ALMA observations of \COthree observed towards R~Scl, and focus on the hydrodynamical modelling of the spiral structure. We here present the full set of \CO~observations of the circumstellar environment around R~Scl, concentrating on the overall CO emission. In Section~\ref{s:observations} we present the ALMA observations of \COone, \COtwo, and \COthree. We also present new SD observations of \COtwo and \COthree taken with the Atacama Pathfinder Experiment telescope (APEX). In Sect.~\ref{s:results} we describe our analysis strategy and the results, and in Sect.~\ref{s:discussion} we discuss what this implies for the physical properties of the gas shell and its origin. Our conclusions are finally presented in Sect.~\ref{s:conclusions}. 
	
	In previous publications, the shells of gas found around carbon AGB stars are referred to as \emph{detached} shells. At least in the case of R~Scl there are clear signs that the shell is in fact not detached; Sects.~\ref{s:ifvssd} and~\ref{s:history}. Throughout the paper we will therefore refer to the \emph{shell} around R~Scl, meaning the thin shell of dust and gas found at 19\farcs5 from the star. We will refer to the gas \emph{between} the shell and the star as the circumstellar envelope (CSE).

%__________________________________________________________________

\section{Observations}
\label{s:observations}

\subsection{R Sculptoris}
\label{s:rscl}

R~Scl is a semi-regular pulsating carbon AGB star with a pulsation period of approximately 370 days. Current distance estimates are very uncertain, ranging from 266 pc~\citep[revised Hipparcos,][]{knappetal2003} to 370 pc~\citep[using K-band Period-Luminosity relationships; ][]{knappetal2003, whitelocketal2008}. Integrating the spectral energy distribution of R~Scl and assuming a distance of 370\,pc gives a luminosity of $L$=5200 \Lsun~-- a reasonable value for carbon AGB stars. We will use 370\,pc throughout this paper, noting the high uncertainty in this value. \cite{maerckeretal2012} use a distance of 290\,pc, adopting a value that was erroneously calculated from a period-luminosity relationship in~\cite{schoieretal2005}. This does not affect the main conclusions in~\cite{maerckeretal2012}, but changes the estimated age of the shell around R~Scl from 1800 years to 2300 years. The estimated mass-loss rates also slightly increase. 

The stellar velocity is determined from molecular line observations to be \vlsrstar=$-19$\,\kms. R~Scl is surrounded by a thin shell of dust and gas. The radius and width of the shell were measured with high precision in images of dust scattered, stellar light, to be 19\farcs5$\pm$0\farcs5 and 2\arcsec$\pm$1\arcsec~\citep{olofssonetal2010,maerckeretal2014}, respectively, and the dust-shell radius and width were found to coincide nearly exactly with the shell of gas observed in \COthree with ALMA~\citep{maerckeretal2014}.

\subsection{ALMA early science observations}
\label{s:almaesobs}

\begin{figure*}
\centering
\includegraphics[width=18cm]{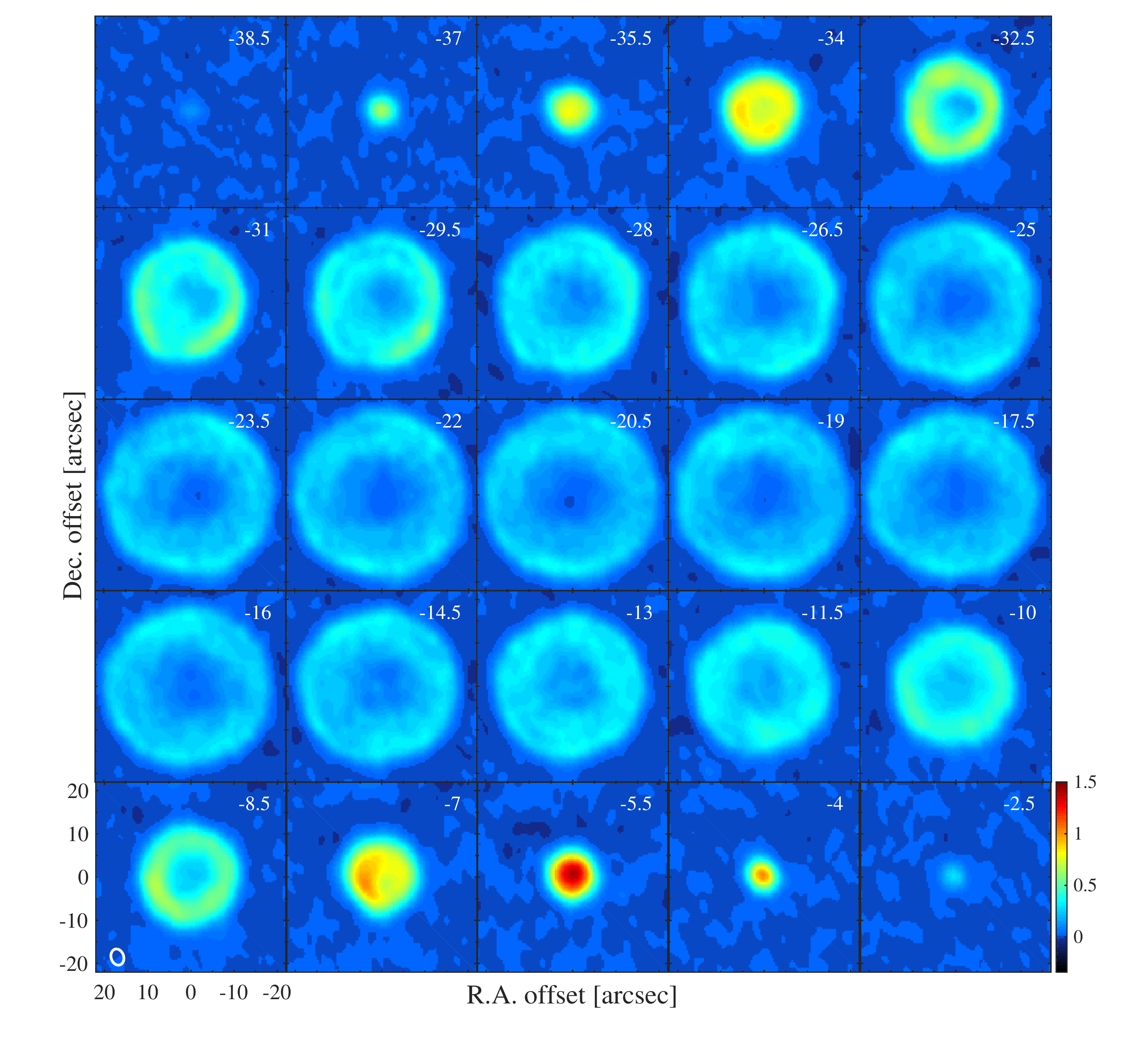}
\caption{ALMA observations of \COone towards R~Scl. The bin size of each panel is 1.5\,\kms. The color scale is given in Jy/beam. The beam ellipse is given in the lower left corner.}
\label{f:co10map}
\end{figure*}

\begin{figure*}
\centering
\includegraphics[width=18cm]{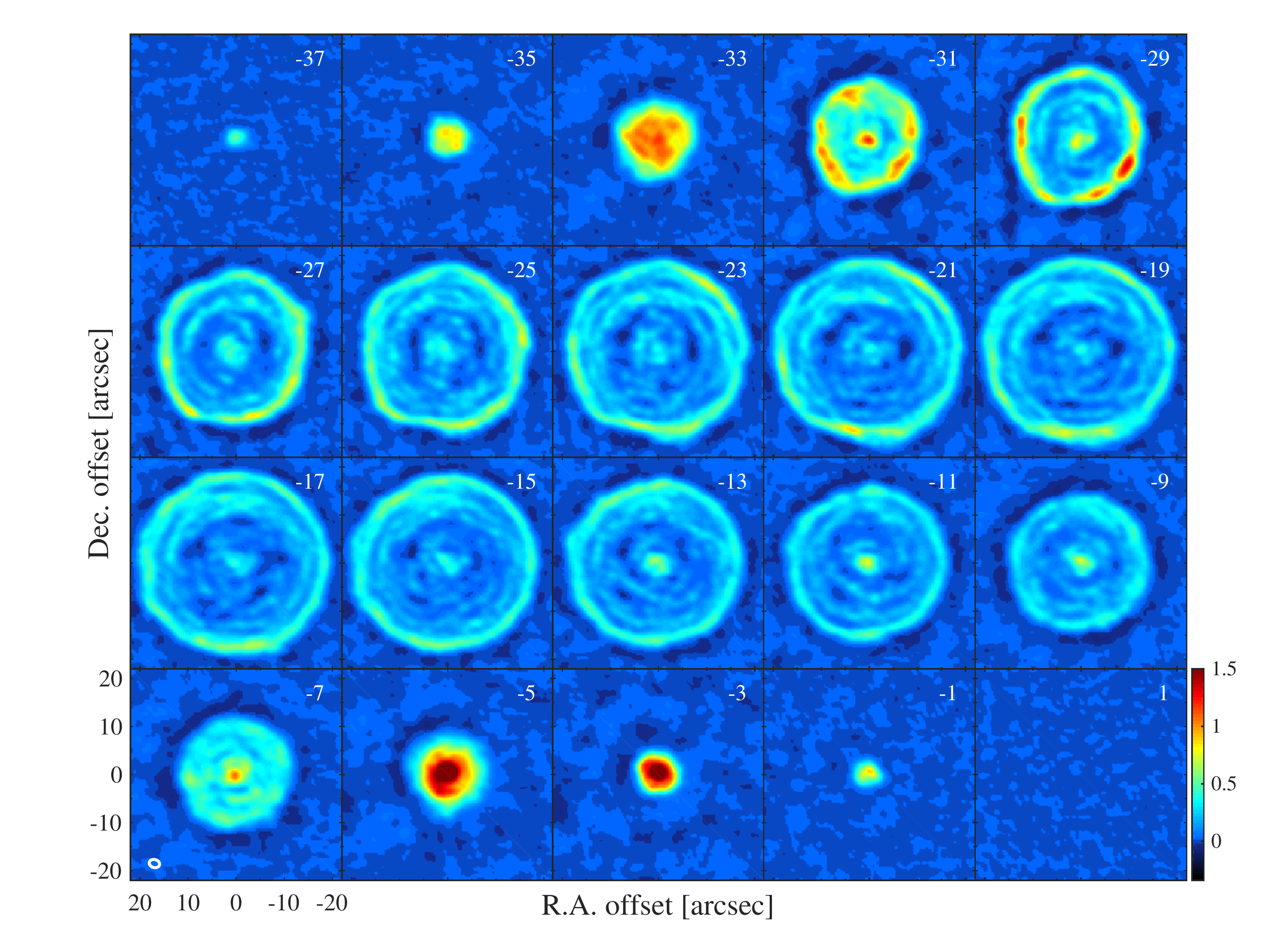}
\caption{ALMA observations of \COtwo towards R~Scl. The bin size of each panel is  1.0\,\kms. Panels with a spacing of 2\,\kms are shown. The color scale is given in Jy/beam. The beam ellipse is given in the lower left corner.}
\label{f:co21map}
\end{figure*}

\begin{figure*}
\centering
\includegraphics[width=18cm]{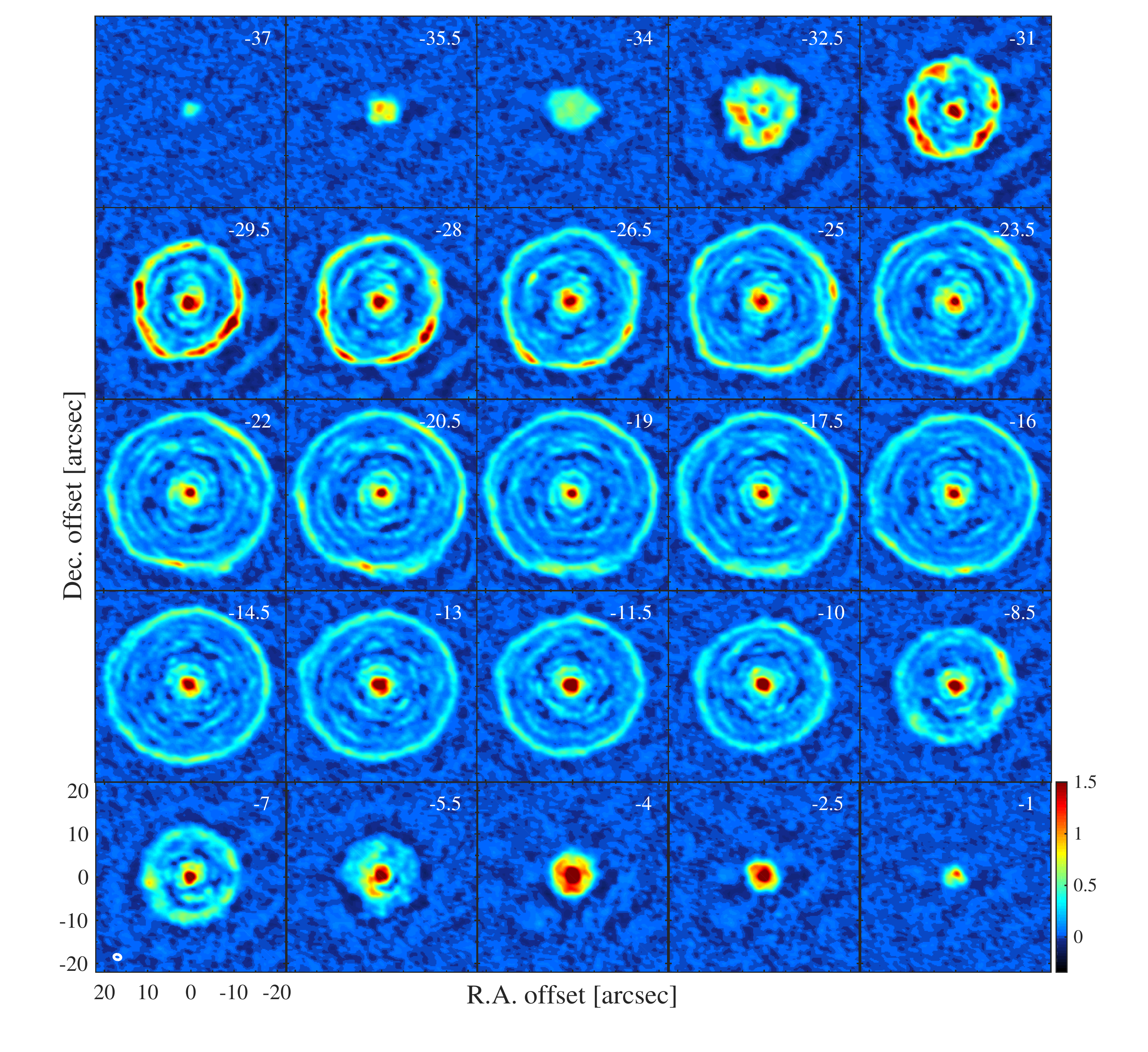}
\caption{ALMA observations of \COthree towards R~Scl. The bin size of each panel is  0.5\,\kms. Panels with a spacing of 1.5\,\kms are shown. The color scale is given in Jy/beam. The beam ellipse is given in the lower left corner. These data were first presented in ~\cite{maerckeretal2012}. }
\label{f:co32map}
\end{figure*}

R~Scl was observed during ALMA Cycle 0 using the compact configuration of the main array. The main target of the observations were the \COone line in Band 3, the \COtwo line in Band 6, and the \COthree line in Band 7. The parameters of the final reduced images are given in Table~\ref{t:coobsim}. Channel maps of the three \CO~transitions are shown in Figs.~\ref{f:co10map} to~\ref{f:co32map}. The observational setup is summarised in Table~\ref{t:coobs} and described below. In Cycle 0 ALMA did not offer the Atacama compact array (ACA) yet, and the data are hence limited by the largest observable scales in the respective transitions. The spectral setup in Cycle 0 allowed us to define (with some restrictions) four spectral windows (spw) with the same bandwidth (BW) and channel width. In all bands pointed mosaics were observed to cover the entire area of the detached shell. The observations are divided into individual scheduling blocks (SBs). The delivered data was re-reduced to improve the data quality using  version 4.1 of the Common Astronomy Software Application (\emph{CASA}), making use of python scripts to automate the calibration process. The SBs are calibrated separately and then combined for the respective bands. After calibration, imaging of the CO emission lines was carried out with the CLEAN algorithm using natural weighting and binning the spectra. The complex structures of the CO envelope were masked in an iterative procedure with decreasing thresholds in CLEAN. 

\noindent
\emph{Band 3:} Observations of the \COone emission line at 115.27\,GHz were carried out between Jan\,23 and Jan\,25 2012. The dataset contains two SBs, taken in the two consecutive nights. Unfortunately, water vapour radiometer (WVR) measurements with sufficient quality were only available for one of the SBs, meaning that rapid atmospheric variations were only corrected for half of the data. One SB revealed very low quality data of the flux calibrator Callisto. Therefore the flux densities of the calibrators were adopted from the other SB. However, the entire spectral window containing the CN(N=1--0) emission had to be excluded from further investigation because of a very low signal-to-noise ratio, making the data unusable.

\noindent
\emph{Band 6:} Observations of the \COtwo emission line at 230.54\,GHz were carried out on Oct 1 2011. The dataset contains two SBs.

\noindent
\emph{Band 7:} Observations of the \COthree emission line at 345.80\,GHz were carried out between Oct\,3 and Oct\,19 2011. The dataset contains a total of seven SBs. One SB did not contain flux calibrator measurements, forcing us to use the flux calibration data from the other SBs.

\begin{table}
\caption{Summary of the reduced image cubes of the ALMA observations of the circumstellar environment of R~Scl. $\theta_{\rm{max}}$ gives the maximum recoverable scale, and the rms is measured in line-free channels in the final reduced images. The absolute calibration uncertainty is $\approx$10-20\%.}
\label{t:coobsim}
\centering
\begin{tabular}{l c c c c}
\hline\hline
Transition		&  Beam & $\theta_{\rm{max}}$ & $\Delta$ v & rms \\
			&  			     &				& [km/s]	& [mJy/beam]\\
\hline
\COone		&  3\farcs8 $\times$ 2\farcs8 & 18\arcsec & 1.5 & 7.3\\
\COtwo		&  2\farcs1 $\times$ 1\farcs7 & 11\arcsec & 1.0 & 6.4\\
\COthree		&  1\farcs6 $\times$ 1\farcs2 & 7\arcsec	  & 0.5 & 16.0\\
\hline\hline
\end{tabular}
\end{table}

\begin{table*}
\caption{Summary of ALMA observations of the circumstellar environment of R~Scl. BW is the total bandwidth, $t_{\rm{tot}}$ the total on-source time in the map, and $\nu_{\rm{spw}}$ the central frequency of the respective spectral windows (with observed lines or continuum indicated). Calibrators are for all spectral windows in the respective bands.}
\label{t:coobs}
\centering
\begin{tabular}{c c c c c c l l l}
ALMA	& BW	& Resolution	& No. of mosaic	& $t_{\rm{tot}}$ 	& \multicolumn{2}{c}{Calibrators} 	& $\nu_{\rm{spw}}$\\
Band		& [MHz]		&	[MHz]	& pointings	&[min]		& bandpass		&	flux		& [GHz]\\
\hline\hline
	3 	& 937.5		& 0.244		& 7			& 64	&	J0522-364& Callisto	&101.5: continuum\\
(84--116\,GHz)&&&&&J040353-360513&	&112.6: CN(N=1--0)\\
&&&&&&	& 113.1: continuum\\
&&&&&&	& 114.5: \COone\\
&&&&&&	& \\
6 &1857.0 &0.488&23&90&3c454.3&Neptune	& 215.25: SiO(5--4) \\
(211--275\,GHz)&&&&&&	& 217.11: continuum\\
&&&&&&	& 230.25: \COtwo\\
&&&&&&	& 232.10: continuum\\
&&&&&&	& \\
7 &1875.0&0.488&45&155&3c454.3&Neptune	&331.17:$^{13}$CO(3--2) \\
(275--373\,GHz)&&&&&J0403-360&	& 333.02: continuum\\
&&&&&&&	 343.30: CS(7--6)\\
&&&&&&&	 345.15: \COthree \\
\hline

\hline\hline
\end{tabular}
\end{table*}

\begin{table}
\caption{Summary of single-dish (SD) observations towards R~Scl. I$_{vlsr}$ gives the measured flux at the stellar \vlsr SD observations. The absolute calibration uncertainty for the SD observations is 20\%. For comparison, the measured flux at the stellar \vlsr measured in the ALMA observations convolved with the respective SD FWHM (Sect.~\ref{s:ifvssd} and Fig.~\ref{f:sdspec}) is also given.}
\label{t:sdobs}
\centering
\begin{tabular}{l c c c c}
\hline\hline
Transition		&  Telescope 	& FWHM 	& I$_{vlsr,SD}$ 	&I$_{vlsr,ALMA}$ \\
			&  			&		& [Jy/beam]	& [Jy/beam]\\
\hline
\COone		&  SEST  		& 44\arcsec & 28 $\pm$1.0&20\\
\COone		&  IRAM  		& 22\arcsec & 11$\pm$1.5&6.5\\
\COtwo		&  APEX 		& 27\arcsec & 75$\pm$1.5&27\\
\COthree		&  APEX 		& 18\arcsec & 50$\pm$2.5&30\\
\hline\hline
\end{tabular}
\end{table}

\subsection{Single-dish observations}
\label{s:sdobs}

In order to estimate the amount of extended emission that goes beyond the largest recoverable scales in the ALMA observations, we compare the ALMA spectra to SD spectra. For the \COone transition we use spectra from the Swedish-ESO submillimeter telescope~\citep[SEST, with a beam full width half maximum (FWHM) of 44\arcsec;][]{olofssonetal1996} and the IRAM 30m telescope~\citep[with a beam FWHM of 22\arcsec;][]{olofssonetal1993a}. 

\begin{figure*}
\centering
\includegraphics[width=18cm]{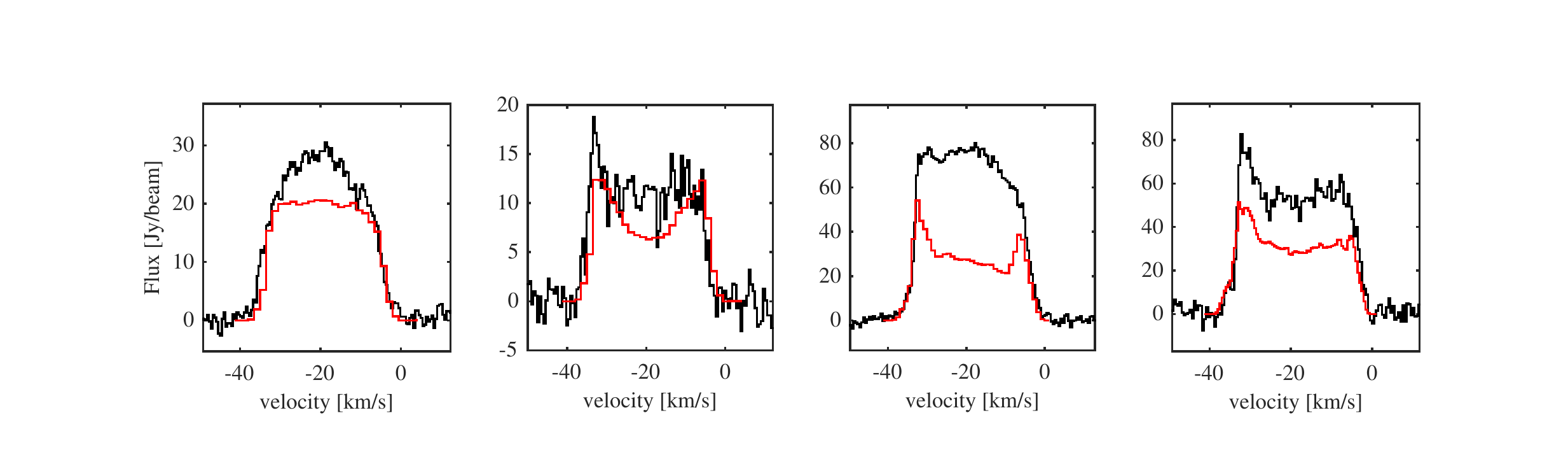}
\caption{SD observations towards R~Scl (from left to right): SEST CO(1--0), IRAM 30m CO(1--0), APEX CO(2--1), and APEX CO(3--2). The FWHM of the SD beams are 44\arcsec, 22\arcsec, 27\arcsec, and 18\arcsec, respectively. The red spectra show the ALMA observations of the respective transitions convolved with Gaussian beams to give the FWHM from the respective telescopes. }
\label{f:sdspec}
\end{figure*}

For the \COtwo and \COthree transitions we observed two on-the-fly (OTF) maps with APEX. The observations were done using the \emph{APEX-1} and \emph{APEX-2} receivers for \COtwo and \COthree, respectively~\citep{vassilevetal2008}.  The beam FWHM is 27\arcsec\/ in the \COtwo transition and 18\arcsec\/ in the \COthree transition. The maps are sampled at $\approx$1/3 of the beam-widths and cover areas of 45\arcsec$\times$45\arcsec\/ and 54\arcsec$\times$54\arcsec, respectively. The total on-source integration of respectively 12.5 min and 13.5 min resulted in rms values of 0.07\,K and 0.18\,K. Figure~\ref{f:sdspec} shows the SD observations. Spectra were extracted from the pixel in the APEX OTF maps centred on the star, giving a spectrum equivalent to a single pointing on the star with the beam of APEX at the respective frequency. The observations are summarised in Table~\ref{t:sdobs}.

\subsection{Interferometer vs. single-dish observations and extended emission}
\label{s:ifvssd}

Figure~\ref{f:sdspec} shows a comparison of observed SD spectra with the equivalent ALMA spectra extracted from maps that were smoothed to the same respective beams using the task \emph{imsmooth} in \emph{CASA}. Depending on the frequency and array configuration, the interferometer will resolve out flux coming from spatial scales that are larger than the largest recoverable scale (Table~\ref{t:coobs}), leading to negative features in the images. The comparison to the SD spectra shows that the ALMA observations miss flux in all the observed emission lines: \\

\noindent
\emph{The front and back of the shell:}\\
One would expect the least amount of flux to be lost at the extreme velocities, where the emission only comes from the front and back caps of the shell and is comparatively compact, while the emission may be more extended at the \vlsrstar due to an extended envelope (see below). For \COone we see that this is indeed the case, where we recover all emission at projected velocities $|\varv_{\rm{LSR}}-\varv_{\rm{LSR}}^*|>11$, while we only recover about 60-70\% of the flux at the $\varv_{\rm{LSR}}^*$. For the \COtwo transition the observations recover $\approx$70\% of the flux at the extreme velocities, while only $\approx$35\% is recovered at the $\varv_{\rm{LSR}}^*$. Finally, 60-70\% of the flux is recovered at all velocities for the \COthree line\footnote{Note that~\cite{vlemmingsetal2013} estimate that only 25\% of the flux is recovered for the \COthree line. This is however due to an erroneous main-beam correction of the APEX data in~\cite{debecketal2010}.}.

At the extreme velocities the geometry of the shell (expansion velocity, shell thickness, and turbulent velocity) causes the caps to have an apparent size that can be larger than the largest recoverable scales. For a shell radius of $R_{\rm{sh}}$=19\farcs5, a shell width of $\Delta R_{\rm{sh}}$=2\arcsec~\citep{maerckeretal2014}, and an expansion velocity of $\varv_{\rm{sh}}$=14.3\,\kms, the size of the cap at the extreme velocity would be $\approx$18\arcsec\/ at 1.5\,\kms resolution in \COone, $\approx$15\arcsec\/ at 1\,\kms resolution in \COtwo, and $\approx$11\arcsec\/ at 0.5\,\kms resolution in \COthree. For \COtwo and \COthree this is indeed larger than the maximum recoverable scale, explaining the lost flux.\\

\noindent
\emph{The CSE and shell:}\\
The amount of flux lost in the different transitions at the $\varv_{\rm{LSR}}^*$ constrains the extended emission of the CSE. In the \COone emission (Fig.~\ref{f:co10map}) the shell can barely be discerned and the spiral shape observed in \COthree~\citep{maerckeretal2012} disappears almost entirely. Instead the ALMA observations show a bowl-like shape with no emission at the stellar position. This can be explained with a shell that is entirely filled with \COone emission, with a low contrast between the extended emission and the small-scale structure (i.e. the shell and spiral). While the SD observations detect all the extended emission inside the shell, the ALMA observations resolve-out the flux at large scales, leading to the observed bowl-like shape. In the \COtwo emission (Fig.~\ref{f:co21map}) the shell is more pronounced and the spiral shape can be seen. However, the structure still lies on top of an overall bowl-like shape, with the present-day mass loss only barely detected. Such an observed intensity distribution can be explained by slightly less extended \COtwo emission, and a higher contrast between the small-scale structures and the extended emission (i.e. less flux gets resolved out compared to the \COone emission, and the small-scale structures are more pronounced). In the \COthree emission (Fig.~\ref{f:co32map}) the shell, spiral, and present-day mass loss are clearly visible, and the lost flux is apparent in the form of negative features around the observed structure. This implies a relatively limited extent of the \COthree emission that is centrally peaked. The maximum recoverable scales for the different transitions in Table~\ref{t:coobs} give an indication of the minimum size of the emitting regions for the individual lines, while the beam FWHMs of the SD spectra give a rough upper limit (depending on whether the line profiles are spatially resolved or unresolved). This confirms that the \COone emission must extend almost out to the shell, causing the smooth distribution and lack of central emission observed in the ALMA data, while the \COtwo emitting region is limited to radii up to $\approx$13\arcsec-15\arcsec, and the \COthree emission comes from a radius up to $\approx$10\arcsec. Extended emission from \CO~reaching out to the shell means that the shell is in fact not detached, but instead filled with significant amounts of molecular gas.

\begin{figure}
\centering
\includegraphics[width=9cm]{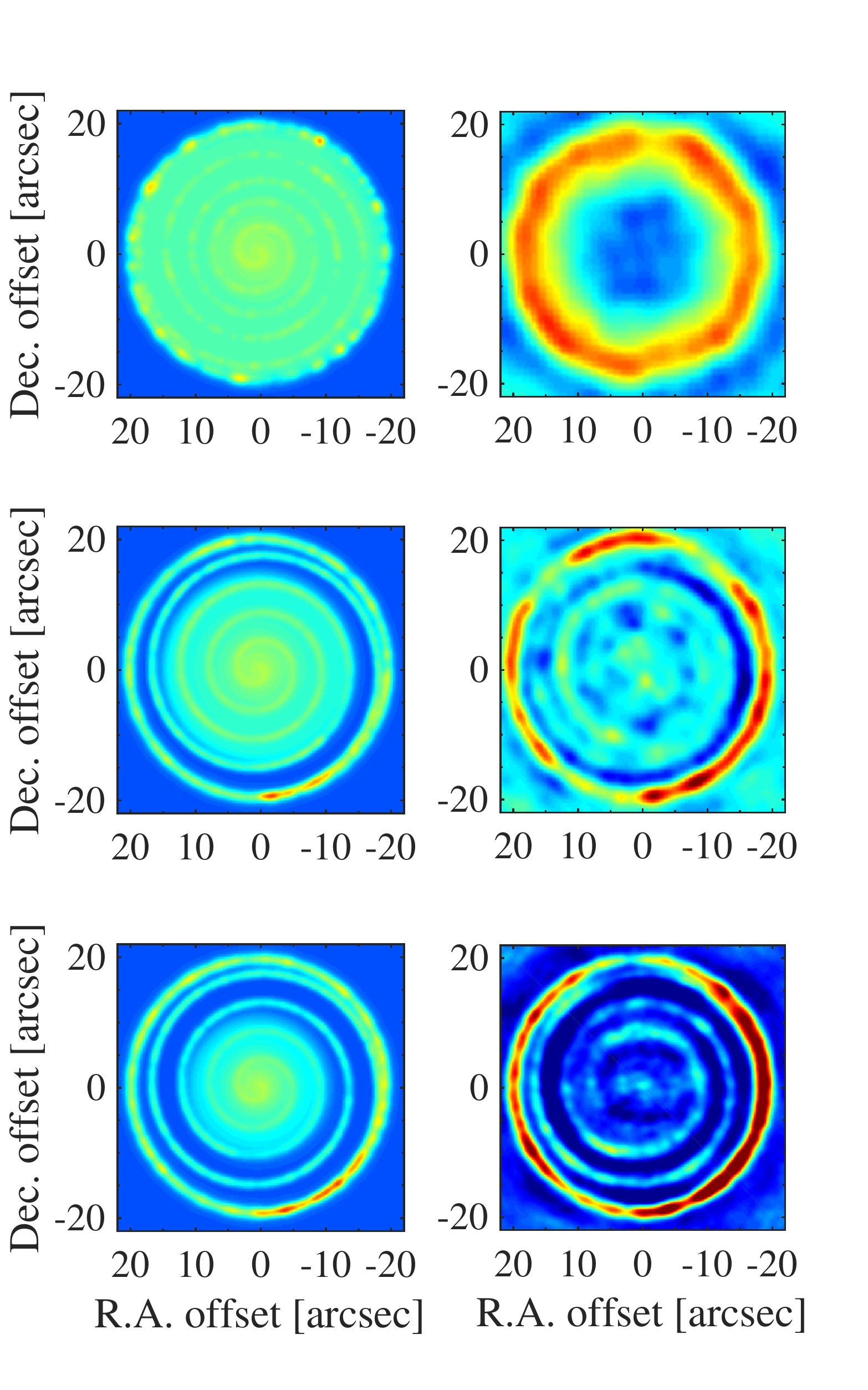}
\caption{Qualitative analysis of the effect of extended emission on the ALMA Cycle 0 observations of ALMA. The left columns shows toy-models of the emission from the shell around R~Scl and extended emission inside the shell at the $\varv_{\rm{LSR}}^*$~for the \COone, \COtwo, and \COthree transitions (top to bottom, respectively). The right column shows the resulting simulated observations using the ALMA Cycle 0 compact configuration of the main array (simulated using the \emph{CASA} simulator).}
\label{f:modsimcomp}
\end{figure}

In Fig.~\ref{f:modsimcomp} we use a toy model to simulate extended regions of emission and their effect on the ALMA observations. We create ad-hoc intensity distributions with no underlying physical model to fit the above described scales and reproduce the observed images. The toy-images consist of a clumpy shell with a radius of 19\farcs5 and a width of 2\arcsec, a spiral with 4.5 evenly spaced spiral windings, and extended emission in the form of discs with radii of $\approx$18\arcsec, $\approx$14\arcsec, and $\approx$9\arcsec\/ for the \COone, \COtwo, and \COthree transitions, respectively, and with a gaussian drop-off at the edge with $\sigma$=6\arcsec.

Simulations in \emph{CASA}, using the ALMA Cycle 0 compact configuration in the three bands, show that the observed features can indeed be explained by a shell and spiral shape where the shell is filled with extended emission of varying size (decreasing from \COone to \COthree). The toy model qualitatively shows the effects of different intensity distributions, explaining the observations, and serves only as a rough estimate of the spatial scales of the emission. The implications this has for the density distribution within the shell, and hence the evolution of the mass-loss rate during and after the TP, are discussed in Section~\ref{s:history}.

%__________________________________________________________________

\section{Analysis and results}
\label{s:results}

We extract the shell emission from the ALMA observations, allowing us to analyse the shell and CSE separately. Our analysis and results focus on the properties of the shell and on the information we can gain about the distribution of the molecular gas of the CSE to constrain the recent mass-loss evolution. We do not treat the spiral. A detailed analysis of the spiral shape observed in \COthree was done using hydrodynamical models~\citep{maerckeretal2012}. Any additional detailed analysis of the CSE structure would require high spatial-resolution observations of the spiral shape to constrain essential parameters such as the binary separation, the shape of the inner spiral, and the contrast between the spiral and inter-spiral material. Such observations are in progress during the current ALMA Cycle (Cycle 2).

\subsection{The shell morphology}
\label{s:shellmorph}

The shell morphology is best described based on the \COthree images, where the ALMA observations give the highest spatial resolution. Overall the shell has a spherical appearance in all velocity channels. Fitting a spherically symmetric shell with a radius of 18\farcs5 expanding at 14.3\,\kms reproduces the \emph{observed} size of the shell in all channels~\citep{maerckeretal2012}. Limb brightening along the inner edge of the shell slightly moves the peak inwards~\citep{maerckeretal2010,olofssonetal2010}, and the radius measured in the ALMA data is consistent with the 19\farcs5 measured in polarised, dust-scattered light of the shell around R~Scl~\citep{maerckeretal2014}. 

However, clear deviations from a perfect sphere are apparent. This was already observed in the dust observations~\citep{maerckeretal2014}, which show an almost identical distribution to the \COthree from ALMA. In particular there is a flattening of the southern part of the shell at the $\varv_{\rm{LSR}}^*$. The shell also appears strongly disrupted in this region. The same structures can also be seen in the \COtwo maps, and to some degree in the \COone maps, where the lower angular resolution, however, smoothes most features out. 

A speculative interpretation of this is an interaction of the binary companion with the stellar wind as the shell is created. The interaction distorts the shell from a spherically symmetric shape, while leaving the shell undisturbed on the opposite side. As the shell expands, the effect of the binary companion is reduced. At the end of the pulse, when the mass-loss rate and expansion velocity start to decline, the spiral formed by the post-pulse mass-loss connects to the shell at position angle PA$\approx$90$^{\circ}$ (with PA=$0^{\circ}$ being North, and the PA increasing counter-clockwise). If true, the flattened and disrupted southern part of the shell would indicate the onset of the TP, while the point at which the spiral attaches to the shell marks the end of the TP. The estimated binary period is based on the present-day expansion velocity of the wind and the separation of the inner spiral windings~\citep{maerckeretal2012} and depends on the estimated distance to the source. Adjusting for the larger distance used here, the binary period is estimated to be 445 years~\citep{maerckeretal2012}. The fraction of the shell affected by the binary companion would be $\approx$40\% of one revolution, translating into a TP timescale of $\approx$180 years.

Whether this is a viable scenario has to be tested through detailed hydrodynamical modelling. The morphology of the shell depends on the mass-loss properties, the masses of the stars, and the timescales. ALMA Cycle 2 data will give us the necessary resolution and sensitivity to constrain these critical parameters and to calculate a detailed model of the shell and spiral formation. 

\subsection{The shell emission}
\label{s:shellemission}

\begin{figure*}
\centering
\includegraphics[width=18cm]{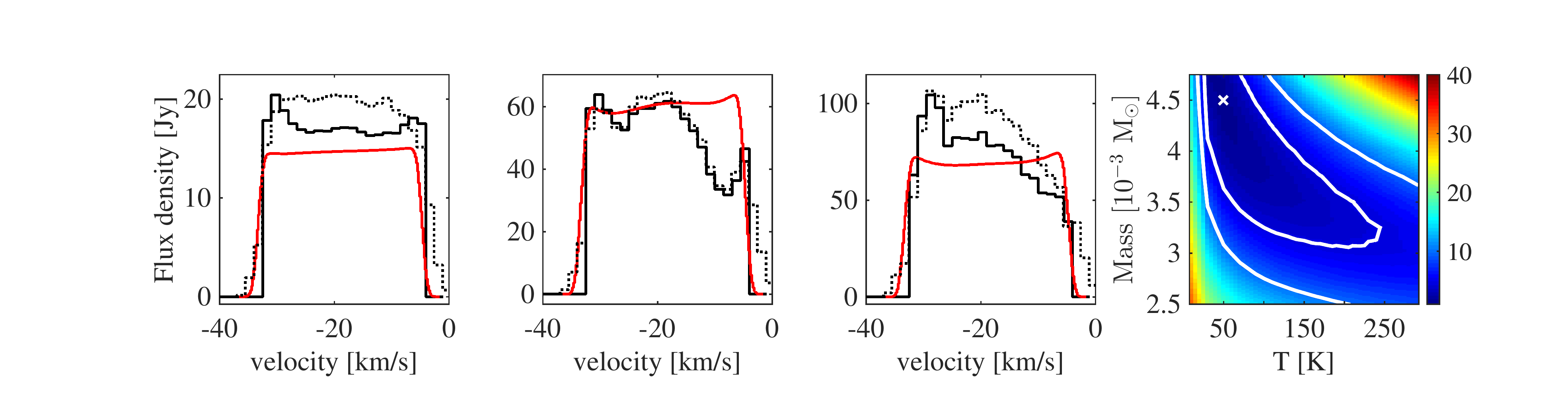}
\caption{ALMA \CO~observations of the shell for \COone, \COtwo, and \COthree (black lines, panels 1 -- 3, respectively). The data are extracted from annuli within $\pm$3\arcsec\/ from the projected radius of the shell in each velocity bin in the images smoothed to 4\arcsec\/ resolution. For comparison, the black dotted lines show the emission integrated over the entire image for each velocity bin instead (i.e. also containing emission from the CSE). The red lines show the best-fit model of the shell only. The right panel shows the $\chi^2$ map of the radiative transfer models of the shell with shell-mass vs. shell temperature. The colour scale gives the $\chi^2$ value, the white cross indicates the best-fit model. The white contours give the 1$\sigma$ and 2$\sigma$ levels (the 3$\sigma$ contours fall outside of the plotted range).}
\label{f:shellspec}
\end{figure*}

In order to analyse the emission from the shell we extract regions in the channel maps for all three transitions assuming a shell-radius of 19\farcs5 and an expansion velocity of 14.3\,\kms~\citep{maerckeretal2012,maerckeretal2014}. In order to ensure that the same regions are extracted in each transition, we smoothed the images to a spatial resolution of 4\arcsec\/ and a velocity resolution of 1.5\,\kms. Although the FWHM of the shell is only 2\arcsec, the spatial and spectral resolution cause a widening of the shell in the images, and the shell emission is extracted from a ring with width $\pm$3\arcsec\/ centred on the shell. The resulting spectra are shown in Fig.~\ref{f:shellspec} and are compared to the spectra integrated over the entire images.

For a homogeneous spherically symmetric shell one would expect a flat spectrum, i.e. each velocity channel probes the same number of CO molecules in the shell. However, in particular for \COtwo and \COthree, it is clear that the shell has a clumpy structure and an asymmetric distribution of CO between the front and the back of the shell. Further, close to the extreme velocities it becomes increasingly difficult to differentiate between shell emission and emission originating from the rest of the CSE. This is particularly true in the case of R~Scl where the emission inside the shell has a velocity gradient that gradually decreases from 14.3\,\kms in the shell to its present-day value of 10.5\,\kms~\citep{maerckeretal2012}.

In order to determine the mass and temperature in the shell, we calculated a grid of 1-dimensional radiative transfer models based on the monte-carlo technique~\citep{schoierco2001}. Each shell has a Gaussian density distribution with a radius of 19\farcs5, a FWHM of 2\arcsec, and a constant expansion velocity of 14.3\,\kms. We assume a fractional abundance of \CO~of f(\CO)=$4.3\times10^{-4}$ relative to H$_2$~\citep{olofssonetal1993a}. This value was derived using the elemental composition given by \cite{lambertetal1986}. The temperature is assumed to be constant throughout the shell. The grid is calculated for shell-masses between $2.5\times10^{-3}$\,\Msun to $4.75\times10^{-3}$\,\Msun in steps of $0.25\times10^{-3}$\,\Msun and temperatures between 10\,K to 290\,K in steps of 20\,K. For each model we determine the best-fit model (Fig.~\ref{f:shellspec}) by minimising the $\chi^2$ value

\begin{equation}
\label{e:chi2}
\chi^2=\sum{\left({I_{\rm{obs,i}}-I_{\rm{mod,i}}}\over{\sigma_{\rm{obs,i}}}\right)^2},
\end{equation}

where $I_{\rm{obs,i}}$ is the integrated intensity of the observation $i$, $I_{\rm{mod,i}}$ is the modelled intensity, $\sigma_{\rm{obs,i}}$ is the uncertainty in the observation is assumed to be 20\%, and the sum goes over all transitions. The best-fit model gives a shell-mass of $(4.5\pm1.5)\times10^{-3}$\,\Msun at a temperature of 50$^{+200}_{-20}$\,K. Note that the radiative transfer model assumes a spherically symmetric, smooth shell, and hence can not reproduce asymmetries within the shell. The observed and modelled line \emph{profiles} may therefore differ significantly. In the optically thin limit, the \emph{integrated intensity} still gives a good measure of the total mass and temperature in the shell. The upper limit in the shell mass is only sampled by few points, and hence not included in the $\chi^2$ map in Fig.~\ref{f:shellspec}. The fractional \CO~abundance further assumes that no \CO~has been photodissociated. Decreasing f(\CO) leads to an increase in the estimated shell mass and a decrease in the shell temperature. For optically thin lines the increase in mass would scale roughly linearly with f(\CO). The lines in our best-fit model have tangential optical depths of $\tau_{\rm{tan}}\sim$1, meaning that the shell mass clearly would be affected by a change in f(\CO), making the estimated shell mass a lower limit. However, for the mass-loss rate during the creation of the shell ($2.3\times10^{-5}$\,\Msunyr, see below), models of photodissociation predict the half-abundance radius of CO to be significantly larger than the radius of the shell~\citep[on the order of 40\arcsec--50\arcsec;][]{staneketal1995}. A clumpy medium will further protect the \CO~in the shell from dissociation (while still allowing photodissociation in the CSE; see Sect.~\ref{s:history}), and we do not expect the photodissociation to affect the abundance of \CO~in the shell significantly. We thus consider the estimated mass to be the correct shell mass within the uncertainties.

The estimate of the mass in the gas shell is consistent with previous estimates from~\cite{schoieretal2005} based on SD data alone ($M_{\rm{shell}}=2.5\times10^{-3}$\,\Msun). More recently, \cite{olofssonetal2015} model the emission of CI from the shell based on SD observations with APEX. Their detection of CI is consistent with the estimated C/O-ratio of R~Scl, assuming solar abundances of oxygen, that all neutral carbon comes from the photodissociation of carbon-bearing molecules except for CO, and there is very limited ionisation of C. They determine the physical parameters of the shell using SD spectra of \CO, including high J-transitions observed with HIFI. Assuming f(\CO)=$10^{-3}$, their resulting shell mass is $(2\pm1)\times10^{-3}$\,\Msun, with a temperature of $100^{+200}_{-70}$\,K. Hence, using essentially independent methods (interferometry vs. SD observations and an independent consistency check with CI observations), and adjusting for the difference in f(\CO), the same shell masses and temperatures are derived.

The analysis of the $^{13}$CO$(3-2)$ emission observed with ALMA suggests that at least parts of the shell must have temperatures that are lower than the derived 50\,K~\citep{vlemmingsetal2013}. The measured $^{12}$CO/$^{13}$CO ratio shows regions in which additional $^{13}$CO is formed due to chemical fractionation. This process however requires temperatures lower than 35\,K. We only model the temperature of a smooth, homogeneous shell, while a clumpy structure will affect the radiative transfer and decouple the connection between mass and temperature.\cite{olofssonetal2015} in principle manage to find a satisfying fit for even lower temperatures than 50\,K, albeit for unrealistically high shell masses. Shell temperatures of $\approx$\,10\,K require masses approaching 1\,\Msun. Although this may in principle be possible, the consistency of a shell mass of a few times $10^{-3}$\Msun with various dust estimates \citep[e.g. ][]{delgadoetal2003a,schoieretal2005,olofssonetal2010} makes a total mass in excess of 0.1\,\Msun unlikely~\citep{olofssonetal2015}.

While the shell is likely confined by collision with a previous, slower wind \citep[e.g.][]{steffenco2000,schoieretal2005,mattssonetal2007}, the mass-loss rate of the previous wind must have been very low (less than a few $10^{-6}$\,\Msunyr). The half-abundance radius due to photodissociation of \CO~lies outside the shell for higher mass-loss rates, but no signs of the spiral structure can be observed outside of the shell.  A significant amount of mass outside of the shell would have further led to a deceleration of the shell, which is not observed~\citep{maerckeretal2012}. Hence, we believe that only a minor amount of mass from the previous mass-loss has been swept up in the shell. Assuming a shell-creation time of $\approx$200$^{+100}_{-50}$ years~\citep[][consistent with the ALMA observations]{vassiliadisco1993} then leads to a mass-loss rate  of $(2.3^{+1.7}_{-1.3})\times10^{-5}$\,\Msunyr during the creation of the shell.

\subsection{The recent evolution of the stellar wind}
\label{s:history}

\begin{figure*}
\centering
\includegraphics[width=18cm]{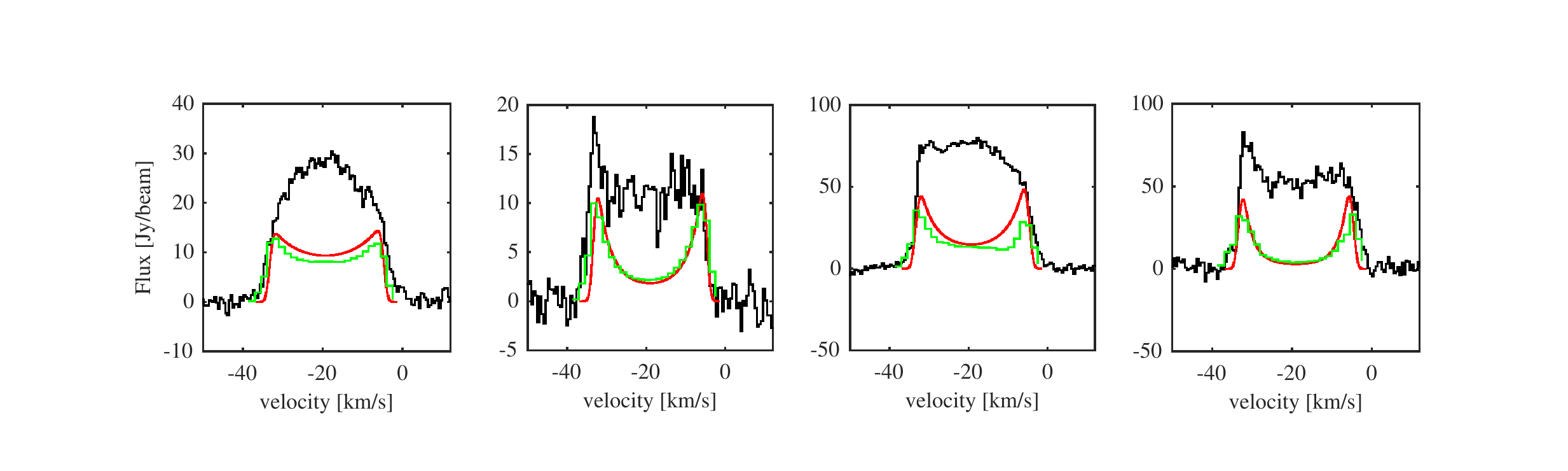}\\
\vspace{-0.5cm}
\includegraphics[width=18cm]{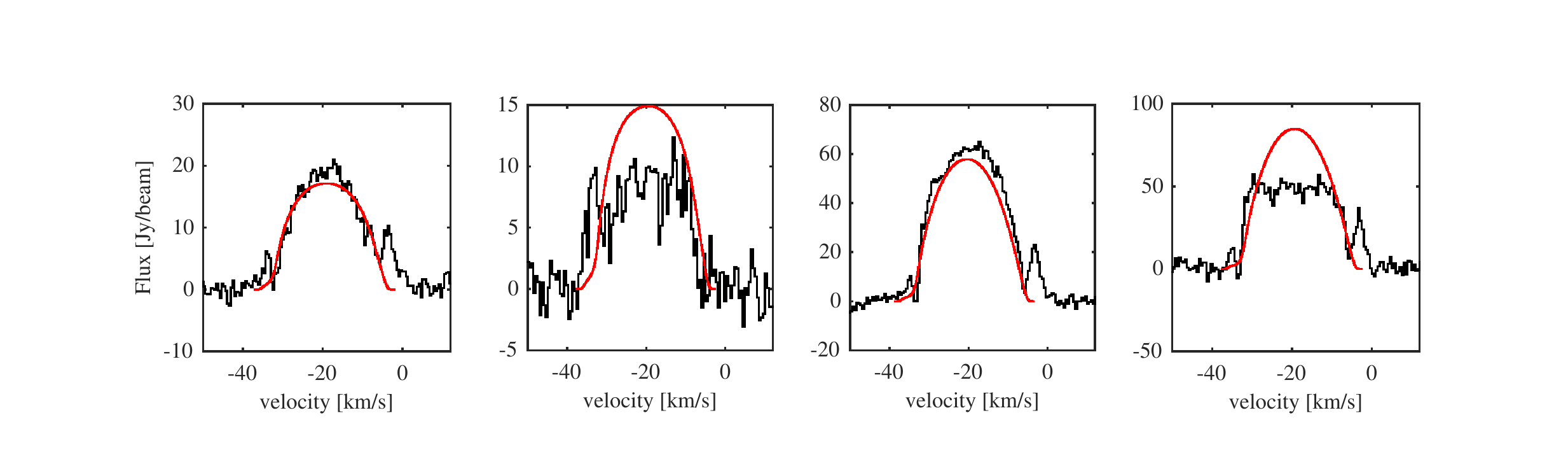}
\caption{\emph{Top:} SD spectra (black, from left to right: SEST~\COone, IRAM~\COone, APEX~\COtwo, and APEX~\COthree) with the best-fit shell model convolved with the respective beams (red). The green spectra show the shell emission extracted from the ALMA data and convolved with the respective SD beams. \emph{Bottom:} SD spectra after subtracting the best-fit shell models. The red lines show the best-fit model fit to the shell-subtracted SD observations to constrain the average mass-loss rate since creation of the shell.}
\label{f:sdsub}
\end{figure*}

In order to constrain the physical parameters of the CSE inside the shell, we convolved the best-fit model of the shell with the SD beams of the respective transitions (Fig.~\ref{f:sdsub}, top), and subtract it from the SD observations. The emission from the shell, extracted from the ALMA images, is additionally convolved with the SD beams and compared to the model to ensure that the right flux is subtracted. We find that the clumpy structure is generally smoothed out significantly in the SD observations, giving a good fit of the model to the observations. Using the model instead of the ALMA observations has the advantage that \emph{only} the shell is subtracted, while the observations may still contain additional velocity components that do not belong to the shell. In difference to the hydrodynamical modelling of the \COthree observations~\citep{maerckeretal2012}, we here focus on the overall extended emission from the CSE, instead of the detailed structure.

The shell-subtracted SD observations (Fig.~\ref{f:sdsub}, bottom) now contain the emission from the CSE without the shell, allowing us to model the mass loss since the creation of the shell. We model the subtracted SD spectra with a single, constant mass-loss rate, while in fact the mass-loss rate likely has varied since formation of the shell. As such we are only modelling an \emph{average} mass-loss rate inside the shell. A varying mass-loss rate will however affect the density distribution and the emission regions of different CO transitions, making it difficult to find a consistent model for all observed lines. Likewise, the half-abundance radius assumes a constant wind and we are only determining an \emph{average} radius for the emitting region. Further, the amount of photodissociation inside the shell is not clear as there likely is significant shielding by the shell itself. Finally, the models are very degenerate between the chosen mass-loss rate, half-abundance radius, temperature profile, and velocity profile. 

Due to these limitations we only derive a model that fits the subtracted SD spectra assuming reasonable parameters for the temperature and velocity profiles, and loosely constraining the radius of the emitting region by the limits given by the observed scales in the ALMA data (see Sect.~\ref{s:ifvssd}). The derived mass-loss rate serves as an indication of the average mass-loss rate since the formation of the shell.

Assuming a constant mass-loss rate results in a 1/$r^2$ density distribution. The fractional abundance is assumed to follow

\begin{equation}
\label{e:fcoprofile}
f(r)=f_0\,{\rm{exp}}\left(-{\rm{ln}}(2)\left({r \over R_{1/2}}\right)^a\right),
\end{equation}

where $f_0$ is the initial fractional abundance, $R_{1/2}$ is the half-abundance radius, and a=2.5~\citep{mamonetal1988}. For the kinetic temperature we assume a power-law of the form $T_{\rm{kin}}(r)=T_0\times(r/R_0)^{\beta}$ with $\beta=-0.85$. R$_0$ is the inner radius of the CSE, assumed to lie at 3 times the stellar radius (assuming a black-body with an effective temperature of 2300\,K and luminosity of 5200\,\Lsun). The velocity increases gradually from 3\,\kms to 14.3\,\kms to take into account both the acceleration of the wind in the inner CSE, as well as the time evolution of the terminal wind velocity since creation of the shell. 

Using the above assumptions, we manage to model the \COone transition observed with SEST and the \COtwo transition observed with APEX using a mass-loss rate of 1.6$\times10^{-5}$\,\Msunyr and a half-abundance radius of 7\farcs2. A \CO~radius that is smaller than the shell indicates that the shell is clumpy, and allows for the penetration of photodissosciating radiation into the CSE. The resulting emitting region for \COone is indeed not very centrally peaked and extends out to $\approx$18\arcsec, while the \COtwo emission is more peaked and only extends out to $\approx$15\arcsec. The \COthree emission is the most compact, extending out to a radius of $\approx$11\arcsec. These results are consistent with the limits set by the ALMA and SD data (Sect.~\ref{s:ifvssd}). While the extent of the emitting region in the different transitions is an effect of the excitation conditions, the best-fit model overestimates the emission in the \COone line observed with IRAM and the \COthree emission observed with APEX, in particular at the $\varv_{\rm{LSR}}^*$. For the \COone transition the beam of the IRAM observations probes a much smaller region than the SEST observations (27\arcsec\/ in diameter vs. 44\arcsec), and hence are dominated by the more recent mass loss from R~Scl. The APEX \COthree observations also are dominated by the recent mass loss, both due to more centrally peaked \COthree emission, as well as due to a smaller beam size (18\arcsec). These two transitions can be modelled with the same half-abundance radius but a mass-loss rate of 3.5$\times10^{-6}$\,\Msunyr, consistent with a continuously declining mass-loss rate since the last TP. The present-day mass-loss rate is estimated to be $\approx$3$\times10^{-7}$\,\Msunyr based on HCN models \citep{schoieretal2005}. Additional photodissociation of molecules in the inner wind due to binary interaction~\citep{vlemmingsetal2013} likely makes this estimate a lower limit, with the true present-day mass-loss rate being closer to $10^{-6}$\,\Msunyr.

Hence, modelling of the shell-subtracted SD data gives a rough picture in which the mass-loss rate decreases gradually by more than an order of magnitude from when the shell was created to today. The mass-loss rate derived by fitting the lines that probe the largest extent of the CSE is only slightly lower than the shell mass-loss rate, implying that the mass loss most likely did not drop very quickly after creation of the shell. In principle it would be possible to attempt to derive a more quantitive evolution of the mass-loss rate since the creation of the shell by modelling a variable mass-loss rate that decreases over time. However, the limits set by the SD spectra are relatively uncertain, and the degeneracies of the model (mass-loss rate, CO photodissociation, temperature profile) are not easily constrained. The loss of extended flux in the current ALMA data prevents using these observations to effectively constrain the mass-loss rate evolution, while the lack of intermediate baselines makes combination with the SD data very uncertain. The ALMA Cycle 2 observations of R~Scl will contain Atacama Compact Array (ACA) observations, together with high-resolution main-array observations, and will make a more detailed description of the spiral around R~Scl and measure the emission on all spatial scales possible. This together with detailed hydrodynamical modelling will set stronger constraints on the recent mass-loss evolution and the creation of the shell around R~Scl.

\subsection{An additional velocity component}
\label{s:highvel}

In the \CO~maps observed with ALMA (Figs.~\ref{f:co10map} to~\ref{f:co32map}) there is clear emission at velocities $\pm18$\,\kms relative to the $\varv_{\rm{LSR}}^*$. The expansion velocity of the shell is 14.3$\pm$0.5\,\kms, while the present-day expansion velocity is constrained to less than 10.5\,\kms by HCN and  high-J \CO~line profiles~\citep{schoieretal2005,olofssonetal2015}. The additional velocity component also becomes evident in the SD spectra after subtracting the shell model (most prominently as a red-shifted peak; Fig.~\ref{f:sdsub}). The ALMA maps show that the emission is centred on the star. However, it is not clear what the origin of this higher-velocity component is. The ratio of the high-velocity components between the observed transitions implies that the emission comes from a warm region close to the star. The ratio between the blue and red-shifted parts of this component further indicates emission from an optically thick outflow. A possible explanation would be the onset of a bipolar outflow. If this were an extended outflow, it would have to be exactly along the line of sight, which is unlikely. A compact bipolar outflow would be possible along any direction as long as it does not extend further than the spatial resolution of the \COthree map, i.e. $\approx$1\farcs4. At a distance of 370\,pc this corresponds to a physical size of $\approx500$\,AU in diameter. An alternative explanation may be a recent increase in expansion velocity of a spherical mass-loss. Whether this comes along with an increase in mass-loss rate is uncertain, but it would imply that the evolution of the post-pulse expansion velocity and mass-loss is not necessarily a smooth function of time. 

It is not clear which (if any) of the above scenarios can explain the additional emission at higher velocities observed in the \CO~lines. The high spatial-resolution data with ALMA from Cycle 2 will resolve the inner CSE of R~Scl with a spatial resolution of 0\farcs2. This will help in identifying the origin of the higher-velocity emission.

%__________________________________________________________________

\section{Discussion}
\label{s:discussion}

For over two decades, the observed, geometrically thin gas shells around a handfull of AGB stars have been connected to TPs. Their formation has been described both theoretically~\citep{steffenetal1998,steffenco2000,mattssonetal2007} and observationally~\citep[e.g.][]{olofssonetal1990,olofssonetal1996,olofssonetal2010,schoieretal2005,maerckeretal2010,maerckeretal2012,maerckeretal2014}. It is generally assumed that the increase in mass-loss rate and expansion velocity during a TP leads to a higher-mass shell interacting with a low-density, slower pre-pulse wind, confining the shell and preventing it from very quickly dispersing due to internal pressure. After the TP, the mass-loss rate and expansion velocity drop drastically, only slowly recovering to values similar to the pre-pulse mass-loss during the remainder of the TP cycle. The sudden drop in mass-loss rate and expansion velocity leads to the shell detaching entirely from the present-day mass loss, leaving an essentially empty shell. The TP cycle starts with a high mass-loss-rate phase which lasts a few hundred years, followed by an inter-pulse time of several ten thousand years, depending on the mass of the star~\citep{karakasco2007}. In order to form a detached shell, a sufficient change between the pre-pulse and pulse mass-loss rates in combination with a change in expansion velocity is required~\citep{mattssonetal2007}. The parameters derived in this paper imply a change in mass-loss rate of more than an order of magnitude, accompanied by a change in expansion velocity. The creation of the shell around R~Scl is hence consistent with formation during a TP.

However, instead of showing a detached shell without (significant) \CO~emission between the shell and the present-day mass loss, the ALMA observations of the shell around R~Scl show a continuous mass-loss rate evolution with a significantly slower decline over time. This leads to a \emph{filled} shell rather than a detached shell. The derived mass of the shell combined with estimated TP timescales implies a relatively high mass-loss rate during creation of the shell. Based on the expansion velocity, the age of the shell is $\approx$2300 years, over which the mass-loss rate has slowly decreased by an order of magnitude. This is consistent with the hydrodynamical modelling of the observed spiral shape seen in the \COthree maps~\citep{maerckeretal2012}. Hence, employing two different and independent methods of analysis (detailed hydrodynamical modelling of the structure seen in one transition vs. radiative transfer of the overall emission in all three transitions), we arrive at the same conclusion. The models form a shell during a TP, with a subsequent linear decline of the mass-loss rate to the present-day value. Assuming a TP mass-loss rate of $2.3\times10^{-5}$\,\Msunyr for 200 years and a linear decline to the present-day value of $10^{-6}$\,\Msunyr over an additional 2100 years, the total mass lost by R~Scl since the onset of the most recent TP is 0.03\,\Msun. If the mass-loss rate instead had evolved according to the classical scenario (i.e. with a sudden drop after the TP), the total mass lost would only have been $\approx$0.007\,\Msun.

As the total amount of mass lost during the TP cycle increases monotonically during AGB evolution, the determined mass can be used to roughly estimate the evolutionary stage of R~Scl, depending on the (unknown) main-sequence mass. For a 3\,\Msun star (main-sequence mass) at solar metallicity typical inter-pulse timescales once the third dredge-up has set in are $\approx$60000 years. Using the values derived for R~Scl, and assuming that the present-day mass-loss rate remains constant for the remainder of the TP cycle at $10^{-6}$\,\Msunyr, this implies that up to 0.09\,\Msun will be lost in total during the most recent TP cycle, placing R~Scl at the end of the TP-AGB~\citep{karakasco2007}. A 4.5\,\Msun star (main-sequence mass) on the other hand has typical inter-pulse timescales of $\approx$10000 years. In this case only a total of 0.04\,\Msun will be lost in the most recent TP cycle, and R~Scl would have only evolved about halfway along the TP-AGB~\citep{karakasco2007}.

Our results show that a significant fraction of the mass during a TP cycle will be lost immediately after the pulse due to a relatively slow decline from the thermal-pulse mass-loss rate. The total amount of mass lost during the TP-cycle limits the lifetime on the AGB and hence the number of TPs a star can experience. The mass of the stellar hydrogen envelope will also have a profound effect on the nucleosynthetic processes inside the star (e.g. hot bottom burning). The evolution of the mass-loss rate throughout the TP-cycle therefore also strongly affects the chemical evolution of the star, and the enrichment of the CSE and ISM. 

%______________________________________________________________

\section{Conclusions}
\label{s:conclusions}

We present the full set of ALMA Cycle 0 observations of \CO~towards the carbon AGB star R~Scl. The observations clearly resolve the shell around R~Scl and reveal a spiral structure induced by a binary companion. The data allow us to separate the shell from the extended emission inside the shell and analyse both components separately. Radiative transfer modelling of the shell and the present-day mass loss constrains the mass in the shell and average mass-loss rate since the formation of the shell. The results are consistent with a shell creation during a TP with a subsequent decline of the mass-loss rate. In particular we find that

\begin{itemize}
\item contrary to what was believed so far, the shell around R~Scl is entirely filled with gas and as such is not detached,
\item the shell is consistent with an increase in mass-loss rate and expansion velocity during a TP, leading to a two-wind interaction with a previous, slower wind,
\item the estimated shell mass-loss rate is $2.3\times10^{-5}$\,\Msunyr for a period of $\approx$200 years, while the post-pulse mass-loss rate is on average $1.6\times10^{-5}$\,\Msunyr during the last $\approx$2100 years, and goes at least as low as  $3.5\times10^{-6}$\,\Msunyr in recent years,
\item the derived mass-loss rates imply that the decline in mass-loss rate after the pulse must have been significantly slower than previously assumed, and
\item the amount of mass lost during a TP cycle may be significantly higher than previously assumed, strongly affecting the lifetime on the AGB, and the chemical evolution of the star, CSE, and ISM.
\end{itemize}

Additionally, we observe a higher-velocity component centred on the star for which we cannot find an obvious explanation. ALMA Cycle 2 observations at high-angular resolution will give a more complete picture of the spiral structure close to the star and include ACA observations to recover also the extended flux. The data will be essential for deriving a more accurate mass-loss rate evolution since the creation of the shell and will allow us to effectively constrain hydrodynamical models. 

\acknowledgements{This paper makes use of ALMA data from project no. ADS/ JAO.ALMA\#2011.0.00131.S. ALMA is a partnership of ESO (representing its member states), the NSF (USA) and NINS (Japan), together with the NRC (Canada) and NSC and ASIAA (Taiwan), in cooperation with the Republic of Chile. The Joint ALMA Observatory is operated by ESO, AUI/NRAO and NAOJ. M.M. has received funding from the People Programme (Marie Curie Actions) of the EU's FP7 (FP7/2007-2013) under REA grant agreement No. 623898.11. W.V. acknowledges support from Marie Curie Career Integration Grant 321691 and ERC consolidator grant 614264. F.K. and M.B. acknowledge funding by the Austrian Science Fund FWF under project number P23586. M.B. further acknowledges funding through the ESO Director General's Discretionary Fund and the uni:docs fellowship of the University of Vienna. HO acknowledges financial support from the Swedish Research Council.}

\bibliographystyle{aa} 
\bibliography{maercker}

\end{document}